\documentclass[prl,twocolumn,preprintnumbers,superscriptaddress,showpacs,amsmath,amssymb]{revtex4}
\usepackage{graphicx,dcolumn,bm}

\begin{document}

\title{Strong valence fluctuation in the quantum critical heavy fermion superconductor
$\beta$-YbAlB$_4$: A hard x-ray photoemission study}

\author{M.~Okawa}
\affiliation{Institute for Solid State Physics, University of Tokyo,
Kashiwa, Chiba 277-8581, Japan}

\author{M.~Matsunami}
\altaffiliation[Present address: ]{UVSOR Facility, Institute for Molecular Science,
Okazaki, Aichi 444-8585, Japan.}
\affiliation{Institute for Solid State Physics, University of Tokyo,
Kashiwa, Chiba 277-8581, Japan}
\affiliation{RIKEN SPring-8 Center, Sayo, Hyogo 679-5148, Japan}

\author{K.~Ishizaka}
\altaffiliation[Present address: ]{Department of Applied Physics, University of Tokyo,
Tokyo 113-8656, Japan.}
\affiliation{Institute for Solid State Physics, University of Tokyo,
Kashiwa, Chiba 277-8581, Japan}

\author{R.~Eguchi}
\altaffiliation[Present address: ]{Graduate School of Natural Science and Technology,
Okayama University, Okayama 700-8530, Japan.}
\affiliation{Institute for Solid State Physics, University of Tokyo,
Kashiwa, Chiba 277-8581, Japan}
\affiliation{RIKEN SPring-8 Center, Sayo, Hyogo 679-5148, Japan}

\author{M.~Taguchi}
\affiliation{RIKEN SPring-8 Center, Sayo, Hyogo 679-5148, Japan}

\author{A.~Chainani}
\affiliation{RIKEN SPring-8 Center, Sayo, Hyogo 679-5148, Japan}

\author{Y.~Takata}
\affiliation{RIKEN SPring-8 Center, Sayo, Hyogo 679-5148, Japan}

\author{M.~Yabashi}
\affiliation{RIKEN SPring-8 Center, Sayo, Hyogo 679-5148, Japan}
\affiliation{Japan Synchrotron Radiation Research Institute, Sayo, Hyogo 679-5198, Japan}

\author{K.~Tamasaku}
\affiliation{RIKEN SPring-8 Center, Sayo, Hyogo 679-5148, Japan}

\author{Y.~Nishino}
\altaffiliation[Present address: ]{Research Institute for Electronic Science,
Hokkaido University, Sapporo 060-0812, Japan.}
\affiliation{RIKEN SPring-8 Center, Sayo, Hyogo 679-5148, Japan}

\author{T.~Ishikawa}
\affiliation{RIKEN SPring-8 Center, Sayo, Hyogo 679-5148, Japan}
\affiliation{Japan Synchrotron Radiation Research Institute, Sayo, Hyogo 679-5198, Japan}

\author{K.~Kuga}
\affiliation{Institute for Solid State Physics, University of Tokyo,
Kashiwa, Chiba 277-8581, Japan}

\author{N.~Horie}
\affiliation{Institute for Solid State Physics, University of Tokyo,
Kashiwa, Chiba 277-8581, Japan}

\author{S.~Nakatsuji}
\affiliation{Institute for Solid State Physics, University of Tokyo,
Kashiwa, Chiba 277-8581, Japan}

\author{S.~Shin}
\affiliation{Institute for Solid State Physics, University of Tokyo,
Kashiwa, Chiba 277-8581, Japan}
\affiliation{RIKEN SPring-8 Center, Sayo, Hyogo 679-5148, Japan}

\date{\today}

\begin{abstract}
Electronic structures of the quantum critical superconductor $\beta$-YbAlB$_4$ and its
polymorph $\alpha$-YbAlB$_4$ are investigated by
using bulk-sensitive hard x-ray photoemission spectroscopy.
From the Yb $3d$ core level spectra, the values of the Yb valence are estimated to
be $\sim$2.73 and $\sim$2.75 for $\alpha$- and $\beta$-YbAlB$_4$, respectively, thus providing
clear evidence for valence fluctuations.
The valence band spectra of these compounds also sho Yb$^{2+}$ peaks
at the Fermi level.
These observations establish an unambiguous case of a strong mixed valence at quantum
criticality for the first time among heavy fermion systems, calling for a novel scheme
for a quantum critical model beyond the conventional Doniach picture in $\beta$-YbAlB$_4$.
\end{abstract}

\pacs{75.30.Mb, 74.70.Tx, 79.60.-i}

\preprint{Journal-ref: Phys.\ Rev.\ Lett.\ \textbf{104}, 247201 (2010)}

\maketitle

Rare-earth compounds possess rich $f$-electron physics such as heavy fermion formation, magnetic transitions,
valence fluctuation phenomena, and quantum criticality
\cite{Hewson93,Varma76,Mathur98,Stewart01,Settai07,Lohneysen07,Monthoux07,Gegenwart08}.
In particular, near the quantum critical point (QCP: a non-thermal-parameter-induced phase
transition point at absolute zero), a quantum fluctuation can lead to unusual behaviors such as non-Fermi
liquid ground state and unconventional superconductivity;
therefore, quantum criticality is one of the most active problems in the study of correlated electrons.
Around the antiferromagnetic QCP, exotic superconductivity is actually found in many Ce-based heavy fermion systems,
which is considered to be mediated by magnetic fluctuation \cite{Mathur98,Settai07,Monthoux07,Gegenwart08}.
Recently, on the other hand, the valence fluctuation-mediated superconductivity around the
valence transition quantum critical end point has also been discussed in CeCu$_2$(Si,Ge)$_2$
\cite{Yuan03,Holmes04,Miyake07}.
Thus, $f$-electron systems exhibit complex exotic superconductivity, providing a variety of
example for the study on other unconventional superconductivity including high-temperature superconductivity.

The magnetic QCP is frequently explained by the Doniach picture \cite{Doniach77},
considering the competition between two different energy scales of
Rudermann--Kittel--Kasuya--Yosida (RKKY) exchange interaction and Kondo screening effect.
The former causes the local moment magnetism of $f$ electrons, while the latter induces the itinerancy of
$f$ electrons accompanying heavy effective mass via the hybridization with the conduction band
($c$--$f$ hybridization).
The $c$--$f$ hybridization involves the valence fluctuation in rare-earth ions, raising the deviation
from the integral valence of 3.
In the case of Yb-based systems, the $4f$-electron configuration fluctuates between $4f^{13}$ (Yb$^{3+}$)
and $4f^{14}$ (Yb$^{2+}$).
The Yb$^{2+}$ component gives the direct measure of the $c$--$f$ hybridization strength,
which can be quantitatively evaluated using high-energy spectroscopic
methods \cite{Yamamoto07,Yamaoka09,Takata07,Sato04,Moreschini07,Suga05,Matsunami08}.
Preceding studies have revealed intriguing aspects of QCP and valence fluctuation in some Yb-based
heavy fermion systems.
In the chemically tuned quantum critical YbCu$_{5-x}$Al$_x$, for example, the Yb valence increases to
3 approaching the antiferromagnetic region. 
Nevertheless, the small deviation from the trivalent configuration is known to remain at the
antiferromagnetic QCP \cite{Yamamoto07,Yamaoka09}.
In such systems, the role of the valence fluctuation in the quantum critical phenomena is under active
discussion \cite{Yamamoto07,Yamaoka09,Knebel06}, which is not considered in the conventional QCP's picture.

A newly discovered polymorph of \textit{Ln}AlB$_4$ system \cite{Macaluso07} has special attraction
in heavy fermions.
$\beta$-YbAlB$_4$ shows non-Fermi liquid behavior under zero-field and ambient pressure,
indicating the immediate existence of a QCP without any external tuning
(i.e., pressure, magnetic field, or chemical doping) \cite{Nakatsuji08}.
Furthermore, this compound shows superconductivity with the transition temperature $T_c$
of 80 mK, which was found for the first time in Yb-based systems \cite{Nakatsuji08,Kuga08}.
In terms of the electron--hole parallelism between Ce$^{3+}$ ($4f^1$; one electron)
and Yb$^{3+}$ ($4f^{13}$; one hole), $\beta$-YbAlB$_4$ is a unique model compound to study
the superconductivity and its relation to the QCP in heavy fermion systems.
In addition, the Curie--Weiss temperature $|\Theta_{\text{CW}}|$ and the crossover temperature $T^{\ast}$
where the hybridized $f$-electron coherence emerges show relatively higher values of order of 100 K,
suggesting that the RKKY and Kondo energy scales compete at a higher energy \cite{Nakatsuji08}.
The susceptibility shows the Curie--Weiss behavior with the effective moment $\mu_{\text{eff}} \sim 2.9\mu_B$/Yb,
and its temperature dependence over the entire $T$ range can be fit by a crystalline electric
field scheme, suggesting a local nature of $4f$ moment \cite{Macaluso07,Nevidomskyy09}.
In contrast, a quantum oscillation study confirmed that heavy $4f$ electrons contribute to the Fermi surface
below large-$T^{\ast}$ \cite{O'Farrell09}.
The origin of the quantum criticality and superconductivity occurring in $\beta$-YbAlB$_4$
is not yet clarified.
It has been recently discussed whether the QCP can be fully explained by the conventional Doniach picture
or not \cite{Nevidomskyy09,Friedemann09}.
The particular quantum critical behavior in $\beta$-YbAlB$_4$ motivates us to reveal the Yb valence
state as an indicator of the Kondo effect's supremacy over the RKKY interaction and to discuss its role
for quantum criticality and superconductivity.

The progress of hard x-ray photoemission spectroscopy (HX-PES) in recent years has enabled us to probe
the bulk-sensitive core level and valence band electronic structure \cite{Takata07}.
A photoelectron with kinetic energies of 5--8 keV attains an escape depth of
50--100 {\AA} from a crystal surface \cite{Kobayashi03}; thus HX-PES can reveal the bulk electronic structure
of $f$-electron systems, which are known to exhibit strong surface effects.
In particular, quantitative estimations of the valence state have been reported by HX-PES
\cite{Sato04,Yamamoto04,Moreschini07,Suga05,Yamasaki07,Matsunami08}.
Since $\beta$-YbAlB$_4$ exhibits a QCP without external parameter tuning, this is an appropriate
compound for a photoemission study related to quantum critical phenomena.

In this Letter, we report the bulk-sensitive HX-PES measurements of $\beta$-YbAlB$_4$ and
its polymorph $\alpha$-YbAlB$_4$.
In contrast to $\beta$-YbAlB$_4$, the ground state of $\alpha$-YbAlB$_4$ is a heavy Fermi liquid
without QCP behavior and superconductivity, though the values of $T^{\ast}$, $\Theta_{\text{CW}}$, and
$\mu_{\text{eff}}$ are very similar \cite{Macaluso07,Horie09}.
In order to understand the essential difference between $\alpha$- and $\beta$-YbAlB$_4$, the measurements of
both polymorphs should be beneficial.
Yb$^{3+}$ and Yb$^{2+}$ components were observed in Yb ${3d}$ spectra for both $\alpha$- and
$\beta$-YbAlB$_4$, which is direct evidence for the valence fluctuation.
The valence band spectrum of $\beta$-YbAlB$_4$ also clearly shows the Yb$^{2+}$ $4f$ peaks
contributing to the electronic state at the Fermi level.
Our results suggest the breakdown of the conventional Doniach picture in $\beta$-YbAlB$_4$.
We discuss the possible relationship between the strong valence fluctuation and
superconductivity at the QCP in $\beta$-YbAlB$_4$.

High-quality single crystals of $\alpha$- and $\beta$-YbAlB$_4$ were grown by the Al-flux
method as described in the literature \cite{Macaluso07}.
HX-PES experiments were performed at the undulator beamline BL29XUL of SPring-8 synchrotron
facility using a Scienta R4000-10kV hemispherical electron spectrometer \cite{Ishikawa05}.
We chose the photon energy of $h\nu = 7.94$ keV.
We obtained clean surfaces of the samples by fracturing \textit{in situ} 
under the base pressure of $\sim$10$^{-8}$ Pa at room temperature.
Data acquisition was carried out at 20 K.
The energy resolution $\Delta E$ was set to $\sim$250 meV.
The Fermi level $E_F$ was determined by the Fermi edge of an evaporated Au film
connected electrically to the sample.

\begin{figure}
\begin{center}
\includegraphics[width=75mm]{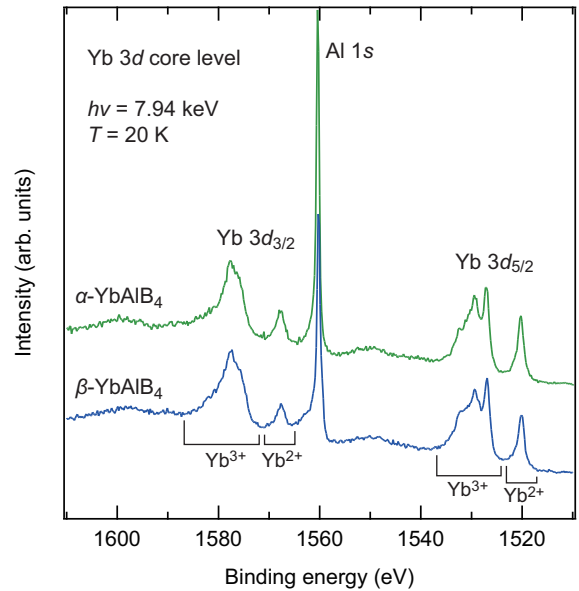}
\end{center}
\caption{HX-PES spectra of Yb $3d$ and Al $1s$ core levels in $\alpha$-YbAlB$_4$ (upper)
and $\beta$-YbAlB$_4$ (lower).}
\label{core}
\end{figure}
Figure \ref{core} shows Yb $3d$ core level spectra of $\alpha$- and $\beta$-YbAlB$_4$.
The obtained spectra are very similar to each other.
The Yb $3d$ level is split into $3d_{5/2}$ and $3d_{3/2}$ levels by the spin--orbit coupling.
The sharp peak at 1560 eV is the Al $1s$ level, while the weak broad peaks at 1550 eV and
1600 eV may be due to plasmon satellites of Yb $3d_{5/2}$ and $3d_{3/2}$ core levels, respectively.
The $3d^{9}4f^{13}$ final state (corresponding to Yb$^{3+}$) multiplets are found at the energy region
of 1524--1538 eV for the $3d_{5/2}$ component and at 1572--1587 eV for the $3d_{3/2}$ component.
The $3d^{9}4f^{14}$ final state (corresponding to Yb$^{2+}$) lines are also observed at 1520 eV and
1567.5 eV as $3d_{5/2}$ and $3d_{3/2}$ components, respectively.
This result, showing the coexistence of the Yb$^{2+}$ and Yb$^{3+}$ levels, is clear evidence
of the valence fluctuation in $\alpha$- and $\beta$-YbAlB$_4$.
Note that all Yb sites in these compounds are crystallographically equivalent \cite{Macaluso07};
therefore, we can exclude the possibility of spatial separation of divalent and trivalent Yb sites.

\begin{figure}
\begin{center}
\includegraphics[width=75mm]{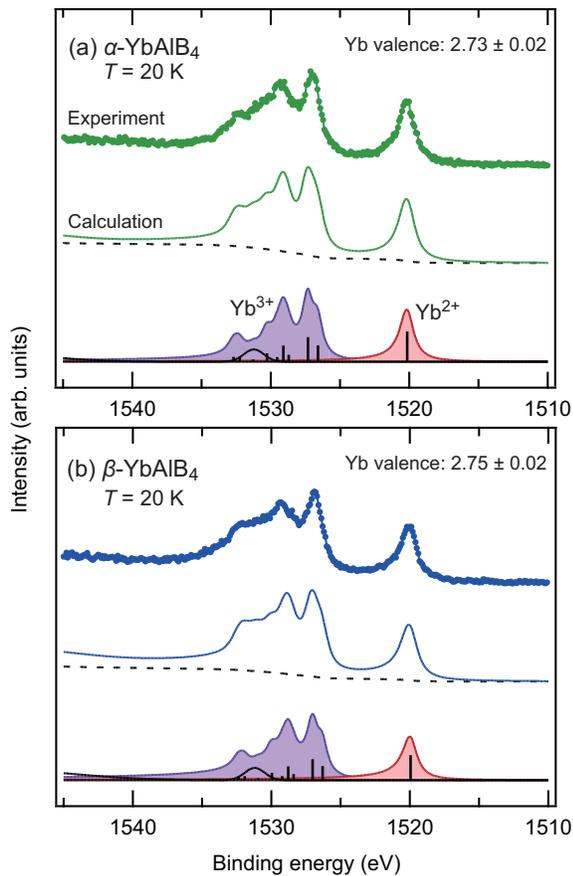}
\end{center}
\caption{Comparison of experimental spectra for Yb $3d_{5/2}$ core level with calculation
in (a) $\alpha$-YbAlB$_4$ and (b) $\beta$-YbAlB$_4$.
Experimental data and fitting results are shown as upper and middle spectra, respectively.
Dashed curves represent the integral background.
Lower spectra represent calculated Yb $3d_{5/2}$ multiplet lines and the broadened (shaded curves) and
plasmon satellites (solid curves).
}
\label{fitting}
\end{figure}

To evaluate the values of Yb valence, we analyzed Yb 3$d_{5/2}$ level spectra which are not
affected by the Al $1s$ peak and its satellite as shown in Fig.\ \ref{fitting}.
Yb$^{3+}$ peaks were reproduced by the result of the atomic multiplet calculation \cite{Cowan81}.
To consider the electron lifetime and the asymmetric spectral shape due to the conduction
electron scattering, each calculated line was broadened using the Doniach--\u{S}unji\'{c}
line shape \cite{Doniach70}.
For the broad plasmon satellites and the background components, we assumed a Gaussian peak
and the Shirley-type integral background \cite{Shirley72}, respectively.
The instrumental energy resolution effect was considered as the Gaussian convolution with the full width of
$\Delta E$.
We can estimate the value of the Yb valence as the ratio of peak intensity areas (shown as shaded areas in
Fig.\ \ref{fitting}) of Yb$^{2+}$ and Yb$^{3+}$ components, thus obtaining $2.73 \pm 0.02$ and
$2.75 \pm 0.02$ as Yb valences for $\alpha$- and $\beta$-YbAlB$_4$, respectively.
These values, significantly deviating from 3, are consistent with high $T^{\ast}$ reflecting
the strong $c$--$f$ hybridization in both compounds.
Considering the similarity in Yb valence, $T^{\ast}$, $\Theta_{\text{CW}}$, and $\mu_{\text{eff}}$,
the ground state of the YbAlB$_4$ system seems sensitive to the difference of the crystal structure.
The strong deviation of Yb valence is also qualitatively consistent with local density
approximation calculations for $\beta$-YbAlB$_4$ (the $4f$-electron number of $n_f \sim 13.4$)
\cite{Nevidomskyy09,O'Farrell09}, though electronic correlations should also be considered to
make more precise comparison.

\begin{figure}
\begin{center}
\includegraphics[width=75mm]{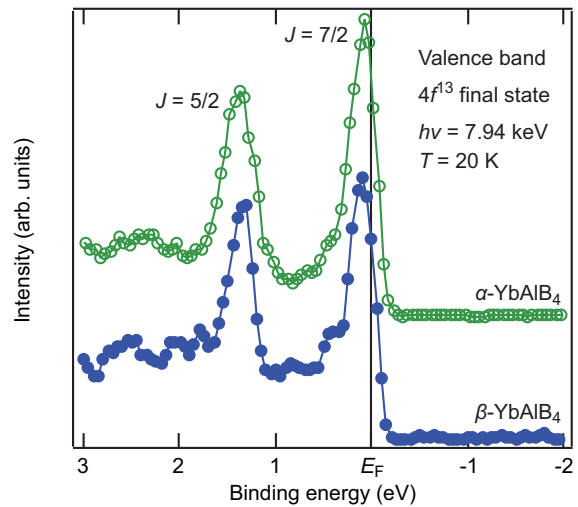}
\end{center}
\caption{HX-PES spectra of the $4f^{13}$ final state in the valence band of $\alpha$- and
$\beta$-YbAlB$_4$.}
\label{valence}
\end{figure}

This valence fluctuation behavior in $\alpha$- and $\beta$-YbAlB$_4$ is also clearly found in the valence band
spectrum shown as Fig.\ \ref{valence}.
The observed peaks at $E_F$ and 1.3 eV correspond to the $J=7/2$ and $5/2$ levels of $4f^{13}$
final state (Yb$^{2+}$) split by the spin--orbit coupling, respectively.
This split width of 1.3 eV is a typical value in Yb-based compounds \cite{Sato04,Suga05}.
The $J=7/2$ level is peaked at $E_F$ and is hybridized with the $s$-$p$ derived conduction band states,
thus acquiring itinerancy, and participating in the Fermi surface.
This indicates that Yb $4f$ electrons become itinerant and its number fluctuates via the $c$--$f$ hybridization,
which is consistent with the core level spectra in Fig.~\ref{core}.

Now we discuss the valence fluctuation behavior in the vicinity of the QCP in $\beta$-YbAlB$_4$.
Since the Yb valence reflects the degree of $4f$-hole localization, the valence fluctuation
suggests a strong hybridization effect of Yb $4f$ holes with the conduction bands for both
$\alpha$- and $\beta$-YbAlB$_4$.
Here, we should note that the valence value of $\sim$2.75 in the non-Fermi liquid $\beta$-YbAlB$_4$
is similar to $\sim$2.73 in $\alpha$-YbAlB$_4$ with normal heavy Fermi liquid behavior \cite{Horie09}.
Indeed, the amount of the divalent component in $\beta$-YbAlB$_4$ is very large
in comparison with other quantum critical Yb-based compounds such as YbCu$_{5-x}$Al$_x$
(Yb$^{\sim 2.95+}$) \cite{Yamamoto07} and YbRh$_2$Si$_2$ (Yb$^{\sim 2.9+}$) \cite{Dallera05}.
We should also note that other comparable valence fluctuation systems such as
YbInCu$_4$ (Yb$^{\sim 2.74+}$) \cite{Sato04} and YbAl$_3$ (Yb$^{\sim 2.71+}$) \cite{Suga05} show
Pauli paramagnetic Fermi liquid behavior at low temperature \cite{Sarrao96,Cornelius02}, indicating
disappearance of the magnetic moment of localized Yb $4f$ electrons due to complete Kondo screening.
Since the magnetic susceptibility of the quantum critical $\beta$-YbAlB$_4$ indicates the existence of the local
moment with Ising anisotropy along the $c$ axis \cite{Nakatsuji08},
such a strong valence fluctuation indicative of the itinerant $4f$ electrons is very unique in
$\beta$-YbAlB$_4$.

This unprecedented dual character of $f$ electrons in $\beta$-YbAlB$_4$ confronts us with difficulty
in the conventional scheme.
In the conventional Doniach phase diagram described by the competition between Kondo screening and RKKY
interaction, the valence fluctuation gets suppressed on approaching QCP.
Here, we conclude that the simple Doniach picture cannot be applied to the quantum critical phenomena
in $\beta$-YbAlB$_4$.
It is worth noting that such strong mixed valence has been unambiguously established for the first
time in heavy fermion compounds, thus showing its possible relation to the unique superconductivity. 
Our finding in $\beta$-YbAlB$_4$ raises the question of how one can microscopically describe the superconducting
mechanism where the $f$ electrons have both itinerant and localized character.
Furthermore, it is an interesting question to check whether $\alpha$-YbAlB$_4$ also shows QCP behavior and the dual
localized-itinerant character, varying an external parameter such as pressure or magnetic field.
To uncover the nature of the QCP and superconductivity in YbAlB$_4$ system, further investigations using detailed
magnetic and theoretical studies across the global phase diagram are highly desired.

In summary, we have investigated the electronic structure in quantum critical $\beta$-YbAlB$_4$
and the Fermi liquid system $\alpha$-YbAlB$_4$ using bulk-sensitive HX-PES with $h\nu = 7.94$ keV.
Both Yb$^{2+}$ and Yb$^{3+}$ components were clearly observed in the Yb $3d$ core level spectra,
providing direct evidence for the noninteger valence values of $\sim$2.73 and $\sim$2.75 for
$\alpha$- and $\beta$-YbAlB$_4$, respectively.
The valence band spectra of these compounds also show the Yb$^{2+}$ peaks at the Fermi level.
We have found the unique valence fluctuation feature at the QCP with the superconductivity in
$\beta$-YbAlB$_4$, thus indicating the possible role of the valence fluctuation in addition
to the magnetic fluctuation, requiring the novel quantum critical scheme beyond the Doniach picture.

The synchrotron radiation experiments (Proposal No.\ 20080068) were performed with
the approval of RIKEN.
This work was supported by Grant-in-Aid for Scientific Research (21684019) from the Japan Society
for the Promotion of Science, by Grants-in-Aid for Scientific Research on Priority Areas (17071003)
and on Innovative Areas ``Heavy Electrons'' (21102507) from the Ministry of Education, Culture, Sports,
Science and Technology (MEXT) of Japan, and by the Kurata Grant.
M.O.\ acknowledges financial support from Global COE Program for ``Physical Science
Frontier,'' MEXT of Japan.


\begin{thebibliography}{37}
\expandafter\ifx\csname natexlab\endcsname\relax\def\natexlab#1{#1}\fi
\expandafter\ifx\csname bibnamefont\endcsname\relax
  \def\bibnamefont#1{#1}\fi
\expandafter\ifx\csname bibfnamefont\endcsname\relax
  \def\bibfnamefont#1{#1}\fi
\expandafter\ifx\csname citenamefont\endcsname\relax
  \def\citenamefont#1{#1}\fi
\expandafter\ifx\csname url\endcsname\relax
  \def\url#1{\texttt{#1}}\fi
\expandafter\ifx\csname urlprefix\endcsname\relax\def\urlprefix{URL }\fi
\providecommand{\bibinfo}[2]{#2}
\providecommand{\eprint}[2][]{\url{#2}}

\bibitem[{\citenamefont{Hewson}(1993)}]{Hewson93}
\bibinfo{author}{\bibfnamefont{A.~C.} \bibnamefont{Hewson}},
  \emph{\bibinfo{title}{The Kondo Problem to Heavy Fermions}}
  (\bibinfo{publisher}{Cambridge University Press},
  \bibinfo{address}{Cambridge}, \bibinfo{year}{1993}).

\bibitem[{\citenamefont{Varma}(1976)}]{Varma76}
\bibinfo{author}{\bibfnamefont{C.~M.} \bibnamefont{Varma}},
  \bibinfo{journal}{Rev. Mod. Phys.} \textbf{\bibinfo{volume}{48}},
  \bibinfo{pages}{219} (\bibinfo{year}{1976}).

\bibitem[{\citenamefont{Mathur et~al.}(1998)\citenamefont{Mathur, Grosche,
  Julian, Walker, Freye, Haselwimmer, Lonzarich}}]{Mathur98}
\bibinfo{author}{\bibfnamefont{N.~D.} \bibnamefont{Mathur}} \textit{et al.},
  \bibinfo{journal}{Nature (London)} \textbf{\bibinfo{volume}{394}},
  \bibinfo{pages}{39} (\bibinfo{year}{1998}).

\bibitem[{\citenamefont{Stewart}(2001)}]{Stewart01}
\bibinfo{author}{\bibfnamefont{G.~R.} \bibnamefont{Stewart}},
  \bibinfo{journal}{Rev. Mod. Phys.} \textbf{\bibinfo{volume}{73}},
  \bibinfo{pages}{797} (\bibinfo{year}{2001}).

\bibitem[{\citenamefont{Settai et~al.}(2007)\citenamefont{Settai, Takeuchi, and
  {\={O}}nuki}}]{Settai07}
\bibinfo{author}{\bibfnamefont{R.}~\bibnamefont{Settai}},
  \bibinfo{author}{\bibfnamefont{T.}~\bibnamefont{Takeuchi}}, \bibnamefont{and}
  \bibinfo{author}{\bibfnamefont{Y.}~\bibnamefont{{\={O}}nuki}},
  \bibinfo{journal}{J. Phys. Soc. Jpn.} \textbf{\bibinfo{volume}{76}},
  \bibinfo{pages}{051003} (\bibinfo{year}{2007}).

\bibitem[{\citenamefont{L{\"o}hneysen et~al.}(2007)\citenamefont{L{\"o}hneysen,
  Bosch, Vojta, and W{\"o}lfle}}]{Lohneysen07}
\bibinfo{author}{\bibfnamefont{H.~v.} \bibnamefont{L{\"o}hneysen}} \textit{et al.},
  \bibinfo{journal}{Rev. Mod. Phys.} \textbf{\bibinfo{volume}{79}},
  \bibinfo{pages}{1015} (\bibinfo{year}{2007}).

\bibitem[{\citenamefont{Monthoux et~al.}(2007)\citenamefont{Monthoux, Pines,
  and Lonzarich}}]{Monthoux07}
\bibinfo{author}{\bibfnamefont{P.}~\bibnamefont{Monthoux}},
  \bibinfo{author}{\bibfnamefont{D.}~\bibnamefont{Pines}}, \bibnamefont{and}
  \bibinfo{author}{\bibfnamefont{G.~G.} \bibnamefont{Lonzarich}},
  \bibinfo{journal}{Nature (London)} \textbf{\bibinfo{volume}{450}},
  \bibinfo{pages}{1177} (\bibinfo{year}{2007}).

\bibitem[{\citenamefont{Gegenwart et~al.}(2008)\citenamefont{Gegenwart, Si, and
  Steglich}}]{Gegenwart08}
\bibinfo{author}{\bibfnamefont{P.}~\bibnamefont{Gegenwart}},
  \bibinfo{author}{\bibfnamefont{Q.}~\bibnamefont{Si}}, \bibnamefont{and}
  \bibinfo{author}{\bibfnamefont{F.}~\bibnamefont{Steglich}},
  \bibinfo{journal}{Nature Phys.} \textbf{\bibinfo{volume}{4}},
  \bibinfo{pages}{186} (\bibinfo{year}{2008}).

\bibitem[{\citenamefont{Yuan et~al.}(2003)\citenamefont{Yuan, Grosche, Deppe,
  Geibel, Sparn, and Steglich}}]{Yuan03}
\bibinfo{author}{\bibfnamefont{H.~Q.} \bibnamefont{Yuan}} \textit{et al.},
  \bibinfo{journal}{Science} \textbf{\bibinfo{volume}{302}},
  \bibinfo{pages}{2104} (\bibinfo{year}{2003}).

\bibitem[{\citenamefont{Holmes et~al.}(2004)\citenamefont{Holmes, Jaccard, and
  Miyake}}]{Holmes04}
\bibinfo{author}{\bibfnamefont{A.~T.} \bibnamefont{Holmes}},
  \bibinfo{author}{\bibfnamefont{D.}~\bibnamefont{Jaccard}}, \bibnamefont{and}
  \bibinfo{author}{\bibfnamefont{K.}~\bibnamefont{Miyake}},
  \bibinfo{journal}{Phys. Rev. B} \textbf{\bibinfo{volume}{69}},
  \bibinfo{pages}{024508} (\bibinfo{year}{2004}).

\bibitem[{\citenamefont{Miyake}(2007)}]{Miyake07}
\bibinfo{author}{\bibfnamefont{K.}~\bibnamefont{Miyake}}, \bibinfo{journal}{J.
  Phys. Condens. Matter} \textbf{\bibinfo{volume}{19}},
  \bibinfo{pages}{125201} (\bibinfo{year}{2007}).

\bibitem[{\citenamefont{Doniach}(1977)}]{Doniach77}
\bibinfo{author}{\bibfnamefont{S.}~\bibnamefont{Doniach}},
  \bibinfo{journal}{Physica B+C (Amsterdam)} \textbf{\bibinfo{volume}{91}},
  \bibinfo{pages}{231} (\bibinfo{year}{1977}).

\bibitem[{\citenamefont{Yamamoto et~al.}(2007)\citenamefont{Yamamoto, Yamaoka,
  Tsujii, Vlaicu, Oohashi, Sakakura, Tochio, Ito, Chainani, and
  Shin}}]{Yamamoto07}
\bibinfo{author}{\bibfnamefont{K.}~\bibnamefont{Yamamoto}} \textit{et al.},
  \bibinfo{journal}{J. Phys. Soc. Jpn.} \textbf{\bibinfo{volume}{76}}, \bibinfo{pages}{124705}
  (\bibinfo{year}{2007}).
  
\bibitem[{\citenamefont{Yamaoka et~al.}(2009)\citenamefont{Yamaoka, Jarrige,
  Tsujii, Hiraoka, Ishii, and Tsuei}}]{Yamaoka09}
  \bibinfo{author}{\bibfnamefont{H.}~\bibnamefont{Yamaoka}} \textit{et al.},
  \bibinfo{journal}{Phys. Rev. B} \textbf{\bibinfo{volume}{80}}, \bibinfo{pages}{035120}
  (\bibinfo{year}{2009}).

\bibitem[{\citenamefont{Takata}(2007)}]{Takata07}
\bibinfo{author}{\bibfnamefont{Y.}~\bibnamefont{Takata}}, in
  \emph{\bibinfo{booktitle}{Very High Resolution Photoelectron Spectroscopy}},
  edited by \bibinfo{editor}{\bibfnamefont{S.}~\bibnamefont{H{\"u}fner}}
  (\bibinfo{publisher}{Springer}, \bibinfo{address}{Berlin},
  \bibinfo{year}{2007}), pp. \bibinfo{pages}{373--397}.

\bibitem[{\citenamefont{Sato et~al.}(2004)\citenamefont{Sato, Shimada, Arita,
  Hiraoka, Kojima, Takeda, Yoshikawa, Sawada, Nakatake, Namatame
  et~al.}}]{Sato04}
\bibinfo{author}{\bibfnamefont{H.}~\bibnamefont{Sato}} \textit{et al.},
  \bibinfo{journal}{Phys. Rev. Lett.} \textbf{\bibinfo{volume}{93}},
  \bibinfo{pages}{246404} (\bibinfo{year}{2004}).

\bibitem[{\citenamefont{Moreschini et~al.}(2007)\citenamefont{Moreschini,
  Dallera, Joyce, Sarrao, Bauer, Fritsch, Bobev, Carpene, Huotari, Vank{\'o}
  et~al.}}]{Moreschini07}
\bibinfo{author}{\bibfnamefont{L.}~\bibnamefont{Moreschini}} \textit{et al.},
  \bibinfo{journal}{Phys. Rev. B} \textbf{\bibinfo{volume}{75}},
  \bibinfo{pages}{035113} (\bibinfo{year}{2007}).

\bibitem[{\citenamefont{Suga et~al.}(2005)\citenamefont{Suga, Sekiyama, Imada,
  Shigemoto, Yamasaki, Tsunekawa, Dallera, Braicovich, Lee, Sakai
  et~al.}}]{Suga05}
\bibinfo{author}{\bibfnamefont{S.}~\bibnamefont{Suga}} \textit{et al.},
  \bibinfo{journal}{J. Phys. Soc. Jpn.} \textbf{\bibinfo{volume}{74}},
  \bibinfo{pages}{2880} (\bibinfo{year}{2005}).

\bibitem[{\citenamefont{Matsunami et~al.}(2008)\citenamefont{Matsunami,
  Chainani, Taguchi, Eguchi, Ishida, Takata, Okamura, Nanba, Yabashi, Tamasaku
  et~al.}}]{Matsunami08}
\bibinfo{author}{\bibfnamefont{M.}~\bibnamefont{Matsunami}} \textit{et al.},
  \bibinfo{journal}{Phys. Rev. B} \textbf{\bibinfo{volume}{78}},
  \bibinfo{pages}{195118} (\bibinfo{year}{2008}).

\bibitem[{\citenamefont{Knebel et~al.}(2006)\citenamefont{Knebel, Boursier,
  Hassinger, Lapertot, Niklowitz, Pourret, Salce, Sanchez, Sheikin, Bonville
  et~al.}}]{Knebel06}
\bibinfo{author}{\bibfnamefont{G.}~\bibnamefont{Knebel}} \textit{et al.},
  \bibinfo{journal}{J. Phys. Soc. Jpn.} \textbf{\bibinfo{volume}{75}},
  \bibinfo{pages}{114709} (\bibinfo{year}{2006});
\bibinfo{author}{\bibfnamefont{G.}~\bibnamefont{Knebel}} \textit{et al.},
  \bibinfo{journal}{Physica (Amsterdam)} \textbf{\bibinfo{volume}{378--380B}},
  \bibinfo{pages}{68} (\bibinfo{year}{2006}).

\bibitem[{\citenamefont{Macaluso et~al.}(2007)\citenamefont{Macaluso,
  Nakatsuji, Kuga, Thomas, Machida, Maeno, Fisk, and Chan}}]{Macaluso07}
\bibinfo{author}{\bibfnamefont{R.~T.} \bibnamefont{Macaluso}} \textit{et al.},
  \bibinfo{journal}{Chem. Mater.} \textbf{\bibinfo{volume}{19}},
  \bibinfo{pages}{1918} (\bibinfo{year}{2007}).

\bibitem[{\citenamefont{Nakatsuji et~al.}(2008)\citenamefont{Nakatsuji, Kuga,
  Machida, Tayama, Sakakibara, Karaki, Ishimoto, Yonezawa, Maeno, Pearson
  et~al.}}]{Nakatsuji08}
\bibinfo{author}{\bibfnamefont{S.}~\bibnamefont{Nakatsuji}} \textit{et al.},
  \bibinfo{journal}{Nature Phys.} \textbf{\bibinfo{volume}{4}},
  \bibinfo{pages}{603} (\bibinfo{year}{2008}).

\bibitem[{\citenamefont{Kuga et~al.}(2008)\citenamefont{Kuga, Karaki,
  Matsumoto, Machida, and Nakatsuji}}]{Kuga08}
\bibinfo{author}{\bibfnamefont{K.}~\bibnamefont{Kuga}} \textit{et al.},
  \bibinfo{journal}{Phys. Rev. Lett.} \textbf{\bibinfo{volume}{101}},
  \bibinfo{eid}{137004} (\bibinfo{year}{2008}).

\bibitem[{\citenamefont{Nevidomskyy and Coleman}(2009)}]{Nevidomskyy09}
\bibinfo{author}{\bibfnamefont{A.~H.} \bibnamefont{Nevidomskyy}}
  \bibnamefont{and} \bibinfo{author}{\bibfnamefont{P.}~\bibnamefont{Coleman}},
  \bibinfo{journal}{Phys. Rev. Lett.} \textbf{\bibinfo{volume}{102}},
  \bibinfo{pages}{077202} (\bibinfo{year}{2009}).

\bibitem[{\citenamefont{O'Farrell et~al.}(2009)\citenamefont{O'Farrell,
  Tompsett, Sebastian, Harrison, Capan, Balicas, Kuga, Matsuo, Kindo, Tokunaga
  et~al.}}]{O'Farrell09}
\bibinfo{author}{\bibfnamefont{E.~C.~T.} \bibnamefont{O'Farrell}} \textit{et al.},
  \bibinfo{journal}{Phys. Rev. Lett.} \textbf{\bibinfo{volume}{102}},
  \bibinfo{eid}{216402} (\bibinfo{year}{2009}).

\bibitem[{\citenamefont{Friedemann et~al.}(2009)}]{Friedemann09}
\bibinfo{author}{\bibfnamefont{S.} \bibnamefont{Friedemann}} \textit{et al.},
  \bibinfo{journal}{Nature Phys.} \textbf{\bibinfo{volume}{5}},
  \bibinfo{pages}{465} (\bibinfo{year}{2009}).

\bibitem[{\citenamefont{Kobayashi et~al.}(2003)\citenamefont{Kobayashi,
  Yabashi, Takata, Tokushima, Shin, Tamasaku, Miwa, Ishikawa, Nohira, Hattori
  et~al.}}]{Kobayashi03}
\bibinfo{author}{\bibfnamefont{K.}~\bibnamefont{Kobayashi}} \textit{et al.},
  \bibinfo{journal}{Appl. Phys. Lett.} \textbf{\bibinfo{volume}{83}},
  \bibinfo{pages}{1005} (\bibinfo{year}{2003}).

\bibitem[{\citenamefont{Yamamoto et~al.}(2004)\citenamefont{Yamamoto, Taguchi,
  Kamakura, Horiba, Takata, Chainani, Shin, Ikenaga, Mimura, Shiga
  et~al.}}]{Yamamoto04}
\bibinfo{author}{\bibfnamefont{K.}~\bibnamefont{Yamamoto}} \textit{et al.},
  \bibinfo{journal}{J. Phys. Soc. Jpn.} \textbf{\bibinfo{volume}{73}},
  \bibinfo{pages}{2616} (\bibinfo{year}{2004}).

\bibitem[{\citenamefont{Yamasaki et~al.}(2007)\citenamefont{Yamasaki, Imada,
  Higashimichi, Fujiwarae, Saita, Miyamachi, Sekiyama, Sugawara, Kikuchi, Sato
  et~al.}}]{Yamasaki07}
\bibinfo{author}{\bibfnamefont{A.}~\bibnamefont{Yamasaki}} \textit{et al.},
  \bibinfo{journal}{Phys. Rev. Lett.} \textbf{\bibinfo{volume}{98}},
  \bibinfo{pages}{156402} (\bibinfo{year}{2007}).

\bibitem[{\citenamefont{Horie et~al.}()\citenamefont{Horie, Tomita, Kuga,
  Matsumoto, Petrovic, and Nakatsuji}}]{Horie09}
\bibinfo{author}{\bibfnamefont{N.}~\bibnamefont{Horie}} \textit{et al.},
  \bibinfo{note}{in \textit{Meeting Abstracts of the Physical Society of Japan}
  (The Physical Society of Japan, Tokyo, 2009), Vol. 64, part 3, p. 651}.

\bibitem[{\citenamefont{Ishikawa et~al.}(2005)\citenamefont{Ishikawa, Tamasaku,
  and Yabashi}}]{Ishikawa05}
\bibinfo{author}{\bibfnamefont{T.}~\bibnamefont{Ishikawa}},
  \bibinfo{author}{\bibfnamefont{K.}~\bibnamefont{Tamasaku}}, \bibnamefont{and}
  \bibinfo{author}{\bibfnamefont{M.}~\bibnamefont{Yabashi}},
  \bibinfo{journal}{Nucl. Instrum. Methods Phys. Res., Sect. A}
  \textbf{\bibinfo{volume}{547}}, \bibinfo{pages}{42} (\bibinfo{year}{2005});
\bibinfo{author}{\bibfnamefont{Y.}~\bibnamefont{Takata}} \textit{et al.},
  \bibinfo{journal}{\textit{ibid.}}
  \textbf{\bibinfo{volume}{547}}, \bibinfo{pages}{50} (\bibinfo{year}{2005}).

\bibitem[{\citenamefont{Cowan}(1981)}]{Cowan81}
\bibinfo{author}{\bibfnamefont{R.~D.} \bibnamefont{Cowan}},
  \emph{\bibinfo{title}{The Theory of Atomic Structure and Spectra}}
  (\bibinfo{publisher}{University of California Press},
  \bibinfo{address}{Berkeley}, \bibinfo{year}{1981}).

\bibitem[{\citenamefont{Doniach and {\u{S}}unji{\'c}}(1970)}]{Doniach70}
\bibinfo{author}{\bibfnamefont{S.}~\bibnamefont{Doniach}} \bibnamefont{and}
  \bibinfo{author}{\bibfnamefont{M.}~\bibnamefont{{\u{S}}unji{\'c}}},
  \bibinfo{journal}{J. Phys. C} \textbf{\bibinfo{volume}{3}},
  \bibinfo{pages}{285} (\bibinfo{year}{1970}).

\bibitem[{\citenamefont{Shirley}(1972)}]{Shirley72}
\bibinfo{author}{\bibfnamefont{D.~A.} \bibnamefont{Shirley}},
  \bibinfo{journal}{Phys. Rev. B} \textbf{\bibinfo{volume}{5}},
  \bibinfo{pages}{4709} (\bibinfo{year}{1972}).

\bibitem[{Dal()}]{Dallera05}
\bibinfo{note}{C. Dallera as discussed in Ref. \onlinecite{Knebel06}}.

\bibitem[{\citenamefont{Sarrao et~al.}(1996)\citenamefont{Sarrao, Immer,
  Benton, Fisk, Lawrence, Mandrus, and Thompson}}]{Sarrao96}
\bibinfo{author}{\bibfnamefont{J.~L.} \bibnamefont{Sarrao}} \textit{et al.},
  \bibinfo{journal}{Phys. Rev. B} \textbf{\bibinfo{volume}{54}},
  \bibinfo{pages}{12 207} (\bibinfo{year}{1996}).

\bibitem[{\citenamefont{Cornelius et~al.}(2002)\citenamefont{Cornelius,
  Lawrencer, Ebihara, Riseborough, Booth, Hundley, Pagliuso, Sarao, Thompson,
  Jung et~al.}}]{Cornelius02}
\bibinfo{author}{\bibfnamefont{A.~L.} \bibnamefont{Cornelius}} \textit{et al.},
  \bibinfo{journal}{Phys. Rev. Lett.} \textbf{\bibinfo{volume}{88}},
  \bibinfo{pages}{117201} (\bibinfo{year}{2002}).

\end{thebibliography}
\end{document}